\documentclass[amsmath,amssymb,aps,pre,superscriptaddress,onecolumn,floatfix,longbibliography]{revtex4-2}
\usepackage{graphicx}
\usepackage{amsmath}
\usepackage{amssymb}
\usepackage{amsfonts}
\usepackage{tabularx}
\usepackage{float}
\usepackage[T1]{fontenc}
\usepackage{bbold}
\usepackage{bm}
\usepackage{hyperref}
\usepackage{subcaption}

\begin{document}

\title{Chaos and anomalous transport in a semiclassical Bose-Hubbard chain}

\author{Dragan Markovi\'c}
\email{vokramnagard@gmail.com}
\affiliation{Department of Physics, University of Belgrade, Studentski Trg 12-16, Belgrade, Serbia}
\affiliation{Center for the Study of Complex Systems, Institute of Physics Belgrade, Pregrevica 118, 11080 Belgrade, Serbia}

\author{Mihailo \v{C}ubrovi\'c}
\email{cubrovic@ipb.ac.rs}
\affiliation{Center for the Study of Complex Systems, Institute of Physics Belgrade, Pregrevica 118, 11080 Belgrade, Serbia}

\date{\today}

\begin{abstract}
    We study chaotic dynamics and anomalous transport in a Bose-Hubbard chain in the semiclassical regime (the limit when the number of particles goes to infinity). We find that the system has mixed phase space with both regular and chaotic dynamics, even for long chains with up to hundred wells. The consequence of the mixed phase space is strongly anomalous diffusion in the space of occupation numbers, with a discrete set of transport exponents. After very long times the system crosses over to the hydrodynamic regime with normal diffusion. Anomalous transport is quite universal, almost completely independent of the parameters of the model (Coulomb interaction, chemical potential): it is mainly determined by the initial distribution of particles along the chain. We corroborate our findings by analytical arguments: scaling analysis for the anomalous regime and the Langevin equation for the normal diffusion regime.
\end{abstract}

\maketitle

\section{\label{secint}Introduction}

The interplay of classical and quantum chaos is an old but still very relevant topic in the dynamics of few- and many-body quantum systems. In recent years there is in particular a surge of interest in quantum chaos, prompted by several novel developments: experimental realization in an increasingly rich array of cold-atom systems \cite{ColdRev}, the appearance of novel theoretical models such as the Sachdev-Ye-Kitaev (SYK) model \cite{SachdevYe,MaldacenaSYK,VandorenSYK}, the appearance of sensitive tools for diagnostics of quantum chaos such as the Krylov complexity \cite{Roberts:2016hpo,Jefferson:2017sdb,Rabinovici:2020,Caputa:2021} and the out-of-time ordered correlators \cite{BlackButter,MSSbound,ScaffidiScrChaos,HashimotoqmOTOC}, and the unexpected connections of quantum chaos to some deep and fundamental problems of high-energy physics such as the black hole scrambling and the information problem of black holes \cite{FastScramblers,BlackButter}. In particular, the Bose-Hubbard model \cite{Bloch:RevModPhys.80.885,Giamarchi:RevModPhys.83.1405,KRUTITSKY20161} is an archetypical strongly correlated quantum-many body system, exhibiting a multitude of interesting phenomena. Unlike the 1D Fermi-Hubbard model, the 1D Bose-Hubbard model is nonintegrable and exhibits quantum chaos \cite{Kolovsky:2004,Kollath:2010,Kolovsky:2016,Kolovsky:2020,Pausch2body}.

The interest for the classical-quantum correspondence in chaotic dynamics is rooted both in reasons of principle (how exactly classical chaos emerges from a quantum chaotic system in the classical limit) and reasons of convenience -- simulations of quantum dynamics are computationally costly thus it is quite valuable if some insight can be gained already from the classical equations of motion. The Bose-Hubbard model is an example of a system where the classical limit exists and can be quite illuminating. Classical dynamics and the classical/quantum correspondence in this model has been the subject of much important work \cite{PolkovnikSachdev2002,Polkovnik2003Main,Polkovnik2003,Graefe:2007,Trimborn:2008,Nakerst:2022prc}. Of particular interest is the fact that the phase space is mixed, i.e., neither regular (the system is nonintegrable) nor fully chaotic. Mixed phase space is a rule rather than exception in few-body systems \cite{Arnold:2013,Lichtenberg:1989} but in \cite{Dag:2022vqb} it was found also in the Bose-Hubbard model where one would not expect it \footnote{The usual expectation is that the existence of a few pseudointegrals of motion or the remnants of KAM tori, which can crucially influence a low-dimensional phase space, would be of little relevance in a phase space of large dimension.}. 

Our goal is to study the influence of weak chaos and mixed regular-chaotic phase space on dynamics and diffusion in a semiclassical one-dimensional Bose-Hubbard chain. While the development and indicators of chaos, both classical \cite{PolkovnikSachdev2002,Polkovnik2003Main} and quantum \cite{Polkovnik2003,Kolovsky:2004,Kolovsky:2007,Graefe:2007,Altman:PhysRevLett.98.180601,Sorg:PhysRevA.90.033606,Haque:PhysRevE.89.042112,Dubertrand_2016,Fischer:PhysRevA.93.043620,Kolovsky:2016,Nakerst:2022prc,PauschSpectrum,PauschOptimalRoute,PauschInterference,Pausch2body} have been studied in great detail in the Bose-Hubbard model, little work exists on the interplay of diffusion and transport in general with the chaoticity of dynamics. It is known that weakly chaotic systems with mixed phase space are the hardest to model statistically, since on one hand the weakness of chaos usually prevents normal diffusion \cite{Lichtenberg:1989} and on the other hand perturbative, purely dynamical treatment is hardly possible for long times whenever the system is nonintegrable. Anomalous scaling and anomalous diffusion are typical for such systems \cite{Zaslavsky1994FractionalKE,ZASLAVSKY2002461,METZLER20001,zaslavsky2007physics} but it is notoriously difficult to classify and explain the anomalous exponents.

In this work we have performed large-scale numerical integrations of classical equations of motion for Bose-Hubbard chains of various lengths (up to 100 sites), finding remarkably simple and robust superdiffusive transport at early times. We will find that populations (ensembles) of initially close orbits spread with time $t$ as $t^{4m}$ or $t^{2m}$ with $m\in\mathbb{N}$ depending on the initial and boundary conditions. Clearly, such fast spread of orbits must end after some time, and after very long times it finally becomes normal diffusion with a linear-in-time growth of the variance. We will also offer some analytical arguments to understand the anomalous regime.

As a final twist, our setup of solving the classical equations of motion for populations of orbits generated from some initial distribution is actually equivalent to the (lowest-order) truncated Wigner approximation (TWA), which corresponds precisely to solving the classical equations of motion with initial conditions averaged over the Wigner quasiprobability function $W$ \cite{PolkovnikAnnals,Polkovnik2018,PolkovnikSciPost}. We can thus interprete our solutions of \emph{classical} equations of motion for \emph{ensembles} as the leading-order TWA approximation for the quantum dynamics, i.e. the semiclassical regime. Such approach to the Bose-Hubbard model has already been explored in \cite{Polkovnik2003}.

The plan of the paper is the following. In Section \ref{secmod} we set up the basics: the classical Hamiltonian and the equations of motion. In Section \ref{secdyn} we study the general dynamical portrait and classical chaos in the system, in particular the mixed nature of the phase space and strong dependence on initial conditions. Section \ref{secdiff} brings our core results: anomalous diffusion of chaotic orbits with a discrete series of scaling exponents, and the transition to normal diffusion at longer times. Section \ref{secconc} sums up the conclusions and the directions of further work.

\section{\label{secmod}Bose-Hubbard model in the classical limit}

\subsection{The classical limit}

We consider 1D Bose-Hubbard model, i.e. a Bose-Hubbard chain with $L$ sites. The bosons are at chemical potential $\mu$, they can hop between neighboring sites with hopping parameter $J$ and at each site interact with Coulomb repulsion $U_\mathrm{BH}$:
\begin{equation}
\label{bh}H_\mathrm{BH}=\sum_{j=1}^L\left[-J\left(b_j^\dagger b_{j+1}+b_jb^\dagger_{j+1}\right)+\frac{U_\mathrm{BH}}{2}n_j\left(n_j-1\right)-\mu n_j\right],
\end{equation}
where $b^\dagger_j,b_j$ are the creation and annihilation operators at the site $j$ and $n_j=b_j^\dagger b_j$ is the number operator. In this work we consider only on-site interaction, i.e. there are no long-range forces. The total number of particles $N\equiv\sum_jn_j$ is an integral of motion as it commutes with $H_\mathrm{BH}$. We are interested in the semiclassical limit, described e.g. in \cite{Polkovnik2003Main,Nakerst:2022prc}. This limit is reached when $N\to\infty$ while the number of sites $L$ stays fixed. Introduction of the complex variables $\psi_j$ as $(b_j^\dagger,b_j)\mapsto (\psi_j^*,\psi_j)\equiv(b_j^\dagger,b_j)/\sqrt{N}$ leads to the vanishing of the commutator: $\left[b_j^\dagger,b_j\right]=1/N\to 0$. This yields the classical Hamiltonian:
\begin{equation}
\label{bhclass}H\equiv\lim_{N\to\infty}\frac{1}{N}H_\mathrm{BH}=\sum_{j=1}^L\left[-J\left(\psi_j^*\psi_{j+1}+\psi_j\psi^*_{j+1}\right)+\frac{U}{2}\vert\psi_j\vert^4-\mu\vert\psi_j\vert^2\right],
\end{equation}
where $U\equiv U_\mathrm{BH}N$ is now the natural measure of the coupling strength (Coulomb repulsion). Following \cite{Nakerst:2022prc}, we can make a useful canonical transformation $\lbrace i\psi_j,\psi_j^*\rbrace\mapsto\lbrace Q_j,P_j\rbrace$ which makes the Hamiltonian more convenient both for analytical and numerical work. The transformation reads
\begin{equation}
\label{pqtrans}Q_j=\frac{i}{\sqrt 2}(\psi_j-\psi^*_j),~P_j=\frac{1}{\sqrt 2}(\psi_j+\psi^*_j).
\end{equation}
In terms of the new variables, the Hamiltonian becomes
\begin{equation}
\label{bhclasspq}H=\sum_{j=1}^L{}'\left[-J\left(Q_jQ_{j+1}+P_jP_{j+1}\right)+\frac{U}{8}\left(Q_j^2+P_j^2\right)^2-\frac{\mu}{2}\left(Q_j^2+P_j^2\right)\right].
\end{equation}
We have denoted the sum by $\sum_{j=1}^L{}'$ because the first and the last term are exceptional (we do not adopt periodic boundary conditions but the usual hard-wall boundary conditions): there are no $P_{j-1},Q_{j-1}$ contributions for $j=1$ and likewise no $P_{j+1},Q_{j+1}$ contributions for $j=L$. The equations of motion read:
\begin{eqnarray}
\nonumber\frac{dP_j}{dt}&=&-J(Q_{j-1}+Q_{j+1})+Q_j\left(\frac{U}{2}\left(P_j^2+Q_j^2\right)-\mu\right)\\
\label{eompq}\frac{dQ_j}{dt}&=&J(P_{j-1}+P_{j+1})-P_j\left(\frac{U}{2}\left(P_j^2+Q_j^2\right)-\mu\right).
\end{eqnarray}
Crucially, the number conservation leads to a constraint which has to be satisfied together with the equations of motion:
\begin{equation}
\label{constraint}\sum_{j=1}^L\vert\psi_j\vert^2=\frac{1}{2}\sum_{j=1}^L\left(P_j^2+Q_j^2\right)=1.
\end{equation}
This system (or equivalently the Hamiltonian (\ref{bhclass})) has been studied extensively \cite{Polkovnik2003Main,Kolovsky:2004,Graefe:2007,Trimborn:2008,Nakerst:2022prc}; it is equivalent to the discrete nonlinear Schr{\"o}dinger equation (DNSE) \cite{Trombettoni:PhysRevLett.86.2353,Kevrekidis:2009}. Being nonintegrable, it is expected to exhibit chaos at least for some initial conditions and parameter regimes; this was indeed found in \cite{Kolovsky:2004} and many subsequent works.

\subsection{Action-angle variables}

We are interested in identifying the relevant dynamical regimes in this system (regular, mixed, chaotic); we want to characterize them quantitatively and then to study the statistical properties of ensembles of orbits, i.e. the distribution functions. The logical path is then to introduce the action-angle variables. According to the well-known paradigm \cite{Arnold:2013,Lichtenberg:1989}, in an integrable system the actions are just the integrals of motion, and in the presence of nonintegrability the Hamiltonian depends also on the angles so the actions evolve in time. But if the chaos is not very strong they are expected to be "slow variables" as opposed to the fast-winding angles. 

Let us now write the Hamiltonian $H$ (Eq.~\ref{bhclasspq}) in terms of action-angle variables. Introducing the actions $I_j$ and their conjugate angles $\phi_j$ as
\begin{equation}
\label{actangle}P_j=\sqrt{2I_j}\sin\phi_j,~~Q_j=\sqrt{2I_j}\cos\phi_j,
\end{equation}
the Hamiltonian becomes:
\begin{equation}
\label{bhclassiphi}H=\sum_{j=1}^L{}'\left(\frac{U}{2}I_j^2-\mu I_j\right)-2J\sum_{j=1}^L{}'\sqrt{I_jI_{j+1}}\cos\left(\phi_j-\phi_{j+1}\right)\equiv H_0\left(\mathbf{I}\right)+JH_1\left(\mathbf{I},\bm{\phi}\right),
\end{equation}
where we have divided the Hamiltonian into the integrable, action-only part $H_0$ and the perturbation $H_1$. The nice thing is that the actions have very simple meaning for the Bose-Hubbard Hamiltonian: from (\ref{actangle}), they are just the occupation numbers (fillings) normalized to unity, i.e. the classical equivalent of the number operators $n_j$ in the quantum Bose-Hubbard Hamiltonian $H_\mathrm{BH}$. Therefore, by studying the system in action-angle variables we simply study the evolution of occupation numbers throughout the chain; from now on, we use the terms action and occupation number as synonymous. The Hamilton equations and the constraint in the new variables read:
\begin{eqnarray}
&&\dot{\phi_j}=-\mu+UI_j-J\left(\sqrt{\frac{I_{j-1}}{I_j}}\cos\left(\phi_j-\phi_{j-1}\right)+\sqrt{\frac{I_{j+1}}{I_j}}\cos\left(\phi_j-\phi_{j+1}\right)\right)\label{eomi}\\
&&\dot{I_j}=2J\left(\sqrt{I_jI_{j-1}}\sin\left(\phi_{j-1}-\phi_j\right)+\sqrt{I_jI_{j+1}}\sin\left(\phi_{j+1}-\phi_j\right)\right)\label{eomphi}\\
&&\sum_{i=j}^LI_j=1\label{constrainti}.
\end{eqnarray}
From (\ref{bhclasspq}) it is obvious that when either $U=0$ or $J=0$ the Hamiltonian reduces to a sum of decoupled sites and the system is integrable. The same is of course true for the form (\ref{bhclassiphi}): for $J=0$ we obviously have $I_j=\mathrm{const.}$, and for $U=0$ we can introduce new variables $(\bar{I}_j,\bar{\phi}_j)$ so that the sites again decouple. The limits of very small or very large $U/J$ ratio will thus exhibit nearly regular dynamics: they correspond to the familiar Mott-insulating and superfluid regimes respectively \cite{PolkovnikSachdev2002,Polkovnik2003Main,Polkovnik2003}.

While the action-angle variables are most convenient for understanding the physics, numerical work is easier in the $(P_j,Q_j)$ variables (Eq.~\ref{pqtrans}) \footnote{Yet another choice of variables in the literature is the one which is manifestly $SU(2L)$ symmetric, thus automatically implementing the constraint (\ref{constraint}) i.e. (\ref{constrainti}), and which follows directly from the basis of coherent states in the full quantum Bose-Hubbard model, as shown in \cite{Trimborn:2008}. While convenient in many situations, such variables are not necessary for our purposes, and we will not use them.}. In Appendix \ref{secappdynnum} we sum up the basic information on numerical integrations in the paper.

\section{\label{secdyn}Classical Bose-Hubbard model: dynamical picture}

\subsection{\label{secdynpor}Phase portrait}

\begin{figure}[H] 
\centering
\includegraphics[width=.9\linewidth]{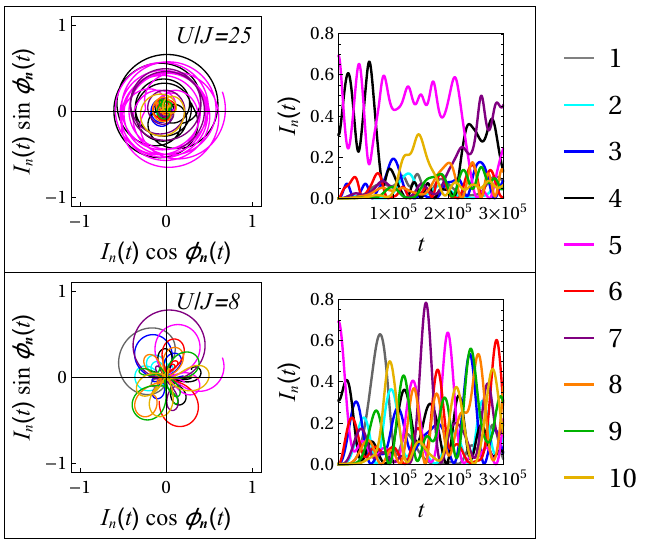}
\caption{Orbits in the action-angle space $I_n(\phi_n)$ (left) and the time evolution of the actions $I_n(t)$ (right) for typical orbits in the Mott-like (top) and superfluid-like (bottom) regimes. The system is the Bose-Hubbard chain at $\mu=0$ with $L=10$ sites, with initial occupations $I_4=0.31,I_5=0.69$ and zero at the remaining sites. The color code stands for the sites 1 thru 10.
}
\label{figorbits}
\end{figure}

As we expect, the system exhibits a typical mixed phase space, with regular islands scattered in the chaotic sea. The obvious control parameter is $U/J$ -- both very small and very large values correspond to near-integrable regimes, whereas in-between we expect strong nonintegrability. Also, at some value $(U/J)_c$ the Mott (localized) regime gives way to the superfluid regime (a priori the relevance of the Mott/superfuid transition for chaos is not clear). These considerations are illustrated by the behavior of typical classical orbits, shown in Fig.~\ref{figorbits}. For large $U/J$, the filled sites initially largely decouple from the others, their motion remains quasiperiodic, and only at later times they transit toward more erratic dynamics. This is the Mott-like regime -- the full sites stay full and the initially localized bosons stay localized for long times. For smaller $U/J$, all orbits mix and all actions show clear secular change.

\begin{figure}[H] 
\centering
\includegraphics[width=.9\linewidth]{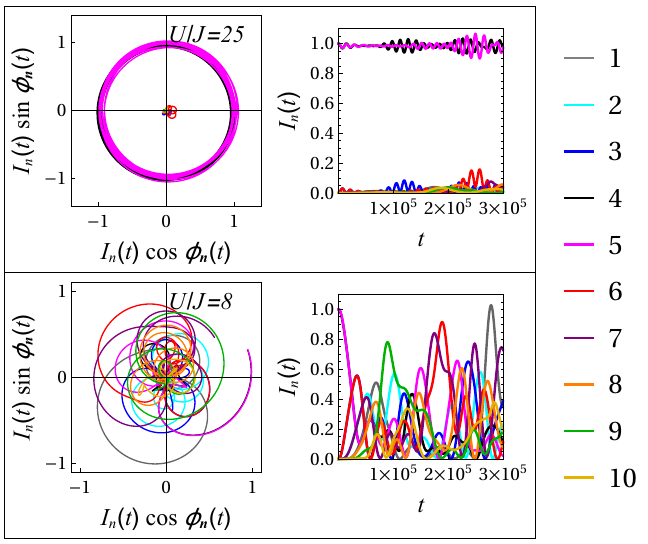}
\caption{Same as in Fig.~\ref{figorbits} but for the resonant initial conditions (Bose-Hubbard chain at $\mu=0$ with $L=10$ sites, with initial occupations $I_4=I_5=1$ and zero at the other sites). For visual convenience the actions in the figure are rescaled by $2$ so the sum of the actions is normalized to $2$ rather than $1$.}
\label{figorbitsres}
\end{figure}

We will see that the perturbative expansion fails in the vicinity of $1:1$ resonances, when two (or more, for resonances of the form $1:1:\ldots :1$) angles have equal frequencies. From the equations of motion (\ref{eomi}-\ref{eomphi}), this happens simply when two neighboring fillings are equal: $I_j=I_{j+1}$. Plotting again the orbits and the evolution of the actions, for the same parameters as in Fig.~\ref{figorbits} but with different initial conditions, starting with a $1:1$ resonance between the occupied sites (site 4 and site 5), Fig.~\ref{figorbitsres} shows that the Mott-like regime (top panel) becomes more persistent and the crossover value of $U/J$ decreases. The stability island of the resonance protects the occupation numbers $I_4$ and $I_5$ from changing much for a long time; the other actions look strongly chaotic, and more so in the superfluid-like regime.

So far we see how the chain moves from the Mott regime to the superfluid regime upon dialing $U/J$, and how the resonant dynamics strengthens the Mott regime. In the subsection \ref{secdynle} we will see how this crosses with the development of chaos.

\subsection{\label{secdynpert}Perturbation theory: non-resonant and resonant}

The equations of motion can be studied within the standard perturbation theory that works as long as we are far enough from low-order resonances \cite{Arnold:2013}. It is natural to treat $J/U$ as a small parameter of nonintegrability as we have done when splitting the Hamiltonian (\ref{bhclassiphi}) into $H_0$ and $JH_1$. For $J=0$ as the zeroth-order approximation we have the integrable system
\begin{equation}
\label{pert0}I_j=I_j^{(0)}=\mathrm{const.},~~\phi_j=\phi_{j0}+\omega_jt=\left(UI_j^{(0)}-\mu\right)t.
\end{equation}
Perturbative corrections are obtained from a series of canonical transformations of the actions, pushing the angle-dependent perturbation at successively higher orders in $t$. The optimal order $\nu$ (which determines the Kolmogorov normal form) is determined by the frequencies of the angles and their proximity to resonances, however for our purposes first-order perturbation theory will be enough to see the hierarchy of dynamical scales and their scaling properties (the key ingredients to understand diffusion).

The first-order correction is obtained from the canonical transformation $\left(\mathbf{I},\bm{\phi}\right)\mapsto\left(\mathbf{I}',\bm{\phi}'\right)$ generated by the function $\chi$ that, for a generic perturbation of the form $H_1=J\sum_{\mathbf{k}}h_{1\mathbf{k}}\exp(-\imath\mathbf{k}\cdot\bm{\phi})$, has the Fourier expansion $\imath J\sum_{\mathbf{k}}\frac{h_{1\mathbf{k}}}{\mathbf{\omega}\cdot\mathbf{k}}\exp(-\imath\mathbf{k}\cdot\bm{\phi})$. In our case the coefficients are nonzero only for wavevectors of the form $\mathbf{k}=(0,\ldots 1,-1,0,\ldots 0)$, so we can keep labelling the sum with the index $i$ instead of the wavevector. For $H_1$ as given in (\ref{bhclassiphi}) the generating function is \footnote{In the following calculations the perturbation is $-J\sum_{j=1}^L{}'\sqrt{I_jI_{j+1}}\cos\left(\phi_j-\phi_{j+1}\right)$ where we have rescaled $2J$ to $J$, for simplicity.}
\begin{equation}
\label{genchi}\chi=\mathbf{I}'\cdot\bm{\phi}+2J\sum_{j=1}^{L}{}'\frac{1}{U}\frac{\sqrt{I'_jI'_{j+1}}}{I'_j-I'_{j+1}}\sin(\phi_j-\phi_{j+1}),
\end{equation}
yielding the first-order corrected Hamiltonian as the sum of the integrable zeroth-order part, the angle-averaged first- and second-order perturbations (also integrable) and the second-order angle-dependent part. Once we perform the above canonical transformation, we use solely the new variables $(\mathbf{I}',\mathbf{\phi}')$, hence we can rename them to $(\mathbf{I},\mathbf{\phi})$ for simplicity, writing the resulting Hamiltonian as:
\begin{eqnarray}
H'(\mathbf{I},\bm{\phi})&=&H_0(\mathbf{I})+J\bar{H}_1(\mathbf{I})+J^2\bar{H}_2(\mathbf{I})+J^2H_2(\mathbf{I},\bm{\phi})=\nonumber\\
&=&\sum_{j=1}^L{}'\left(\frac{U}{2}I_j^2-\mu I_j+\frac{4J^2}{U}\frac{I_jI_{j+1}}{\left(I_j-I_{j+1}\right)^2}\right)+\frac{8J^2}{U}+\nonumber\\
&+&\sum_{j=1}^L{}'\left(h^{(2)}_j\left(\mathbf{I}\right)\cos\left(\phi_j-\phi_{j+1}\right)+\tilde{h}^{(2)}_j\left(\mathbf{I}\right)\cos\left(\phi_j-2\phi_{j+1}+\phi_{j+2}\right)\right).\label{hpert}
\end{eqnarray}
For completeness we have included also the constant term $8J^2/U$ even though it is irrelevant for the dynamics (it would matter for the partition functions and thermodynamics of the system). The expressions for the coefficients of the second-order perturbation $h_{2j},\tilde{h}_{2j}$ can be found explicitly:
\begin{eqnarray}
h^{(2)}_j&=&-\frac{2J^2}{U}\frac{I_jI_{j+1}}{\left(I_j-I_{j+1}\right)^2}\label{h2}\\
\tilde{h}^{(2)}_j&=&-\frac{2J^2}{U}\frac{I_j\sqrt{I_{j-1}I_{j+1}}}{\left(I_j-I_{j-1}\right)^2\left(I_j-I_{j+1}\right)^2}\left[I_{j-1}^2+2I_j^2+I_{j+1}^2-2(I_{j-1}+I_{j+1})I_j\right].\label{h2tilde}
\end{eqnarray}
As could be expected, the perturbation theory fails near $1:1$ resonances $I_j=I_{j\pm 1}$, when the coefficients diverge \footnote{It is reasonable to ask also what happens near other resonances, because we know that any resonance has a region of phase space where the naive perturbation theory fails. However, at first order, only the $1:1$ resonance requires special variables, as we can see from Eqs.~(\ref{h2}-\ref{h2tilde}). We did not explore higher-order perturbation theory, and our numerical results on transport also do not show any special phenomena related to higher-order resonances.}. In order to address resonant dynamics near the $I_j=I_{j+1}$ resonance, we perform the textbook transformation
\begin{equation}
\left(I_j,I_{j+1};\phi_j,\phi_{j+1}\right)\mapsto\left(I_r,I_0;\Phi,\phi\right)=\left(I_j,I_j+I_{j+1};\phi_j-\phi_{j+1},\phi_{j+1}\right),
\end{equation}
finding that $I_0\approx\mathrm{const.}$ so that the effective resonant Hamiltonian only depends on $(I_r,\Phi)$:
\begin{equation}
H_\mathrm{res}=UI_r^2+UI_0I_r-2J\sqrt{I_r(I_0-I_r)}\cos\Phi.\label{hpertres}
\end{equation}
Obviously, for long chains with $L\gg 1$ the structure of the phase space can be extremely complicated, with numerous resonances and both chaotic and regular areas. The scaling properties of the coefficients $h^{(2)}_j,\tilde{h}^{(2)}_j$ in the nonresonant case (Eq.~\ref{hpert}) and in the resonant one (Eq.~\ref{hpertres}) will be useful later when we try to understand the exponents of anomalous transport.

\subsection{\label{secdynle}Lyapunov exponents}

\begin{figure}[H] 
\centering
\includegraphics[width=.95\linewidth]{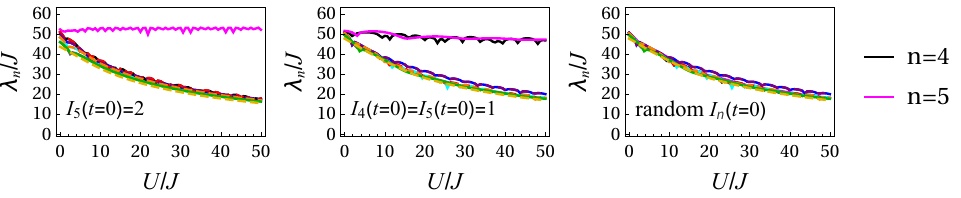}
\caption{The Lyapunov exponent $\lambda_n$ (in units of $J$) as a function of the Coulomb interaction $U/J$ at the chemical potential $\mu/J=0.2$, for a Bose-Hubbard chain with $L=10$ wells, with different initial conditions: single occupied site at $n=5$ (a), two occupied neighboring sites at $n=4$ and $n=5$ i.e., a classical resonance (b), a random chain with arbitrary initial actions/occupancies generated from a uniform distribution (c). The color code (same as in Figs.~\ref{figorbits} and \ref{figorbitsres}) stands for the site number, i.e. the degree of freedom $(I_n,\phi_n)$ for which we compute the Lyapunov exponent; in order not to clutter the figure too much, we have only emphasized the colors for the initially filled sites ($n=4$ and $n=5$), which consistently show stronger chaos. Chaos seems to be driven by the initial conditions, i.e. initial values of the actions $I_n(t=0)$. For better visibility some of the lines are dashed; the dashing does not signify any specific physical property.}
\label{figtable}
\end{figure}

We now probe the classical chaos of various configurations by computing the Lyapunov exponents of the classical equations of motion. There is one Lyapunov exponent per degree of freedom, thus we can label the as exponents $\lambda_n$ ($n=1,\ldots L$), just like the actions. Calculating the Lyapunov exponents for the same configurations as in Figs.~\ref{figorbits} and \ref{figorbitsres} we find that chaos is uniformly strong for filled sites, whereas for initially empty sites it goes down as the ratio $U/J$ increases and the system is closer and closer to the Mott regime (Fig.~\ref{figtable}). In the same figure we show also the case with randomly chosen actions $I_n(t=0)$: this case displays uniformly weak chaos \footnote{Here we tacitly assume from the beginning that higher Lyapunov exponents mean stronger chaos. This is indeed generally true in systems with finite phase space volume; it is not in general true for systems like inverse harmonic oscillator where orbits can escape to infinity. But the latter does not happen in the Bose-Hubbard model so for us indeed the exponents $\lambda_n$ should be a meaningful indicator of chaos.}. The message is thus the following: chaos is not primarily related to localized/superfluid regimes, nor even to the existence of the resonances (in Fig.~\ref{figtable} there cannot be any resonances in the left panel, where only one site is initially filled, yet the phenomenology is the same as in the central panel, with the $1:1$ resonance between $I_4$ and $I_5$). Chaos is in fact determined by the action value: the fuller the site, the stronger the chaos.

\begin{figure}[H] 
\centering
\includegraphics[width=.9\linewidth]{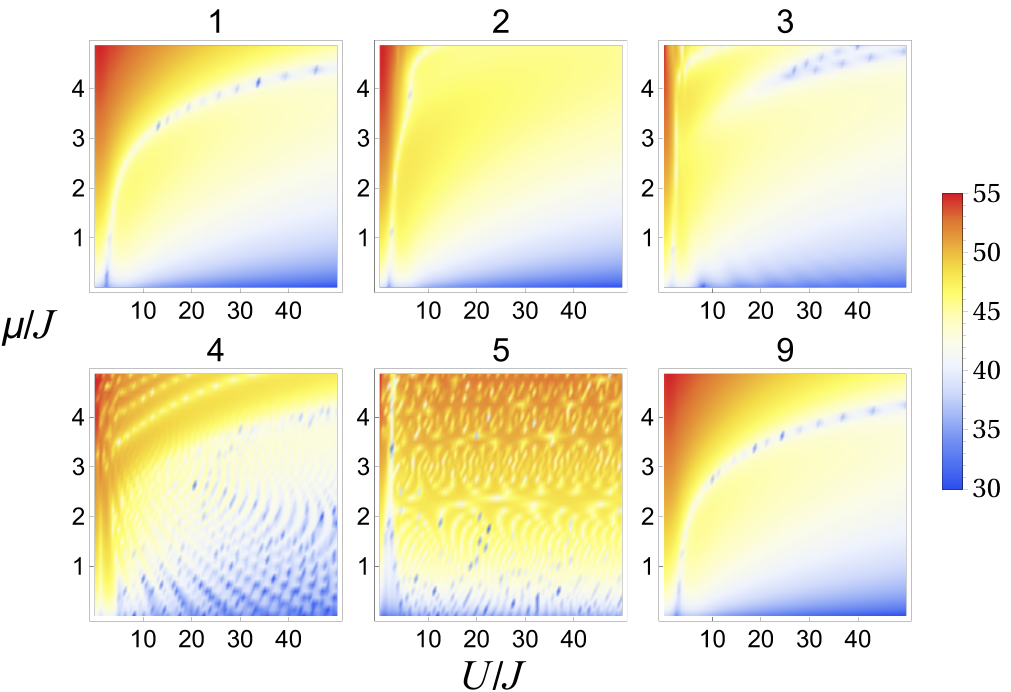}
\caption{The Lyapunov exponent $\lambda$ (in units of $J$) as a function of the Coulomb interaction $U/J$ and the chemical potential $\mu/J$, for a Bose-Hubbard chain with $L=10$ wells, with the well $n=5$ initially full and the rest empty. The Coulomb coupling $U/J$ is varied from $0.5$ to $50.0$, i.e. we do not include the cases with $U/J=0$ for these have exactly zero Lyapunov exponent.}
\label{figtabtablesingle}
\end{figure}

\begin{figure}[H] 
\centering
\includegraphics[width=.9\linewidth]{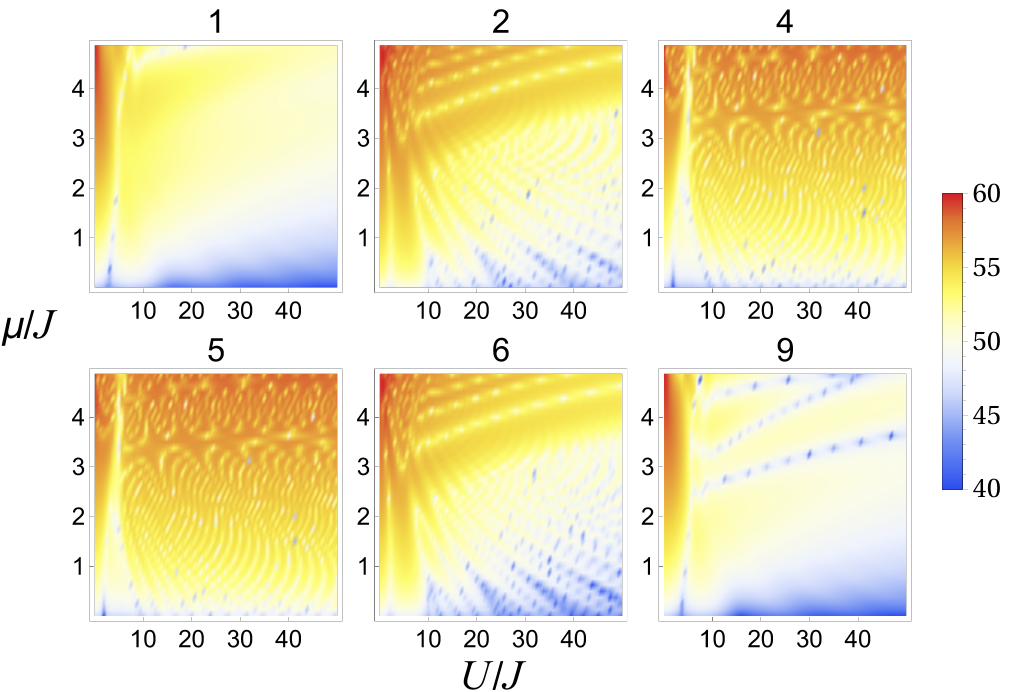}
\caption{The Lyapunov exponent $\lambda$ (in units of $J$) as a function of the Coulomb interaction $U/J$ and the chemical potential $\mu/J$, for a Bose-Hubbard chain with $L=10$ wells, with the wells $n=4$ and $n=5$ initially full and the rest empty. The Coulomb coupling $U/J$ is varied from $0.5$ to $50.0$, i.e. we do not include the cases with $U/J=0$ for these have exactly zero Lyapunov exponent.}
\label{figtabtabledouble}
\end{figure}

This is further corroborated by scanning the values of $\lambda_n$ as a function of both $\mu$ and $U$. Fig.~\ref{figtabtablesingle} shows the value of the Lyapunov exponent $\lambda_n(U/J,\mu/J)$ for different sites in the chain with initially filled site $4$. Although some corners of the parameter space show strong chaos even for sites far from the filled site, the exponent is overall the highest for the initially filled site, and the size of the chaotic region generally \emph{decreases with the distance from the full site.} For the chain with the initial condition $I_4(t=0)=I_5(t=0)=1/2$, the other actions being zero (Fig.~\ref{figtabtabledouble}), $\lambda_4$ and $\lambda_5$ are similarly much larger than the other exponents.

The figures suggest that the initial conditions and specifically the distance from the nearest site with nonzero initial occupation (the distance is zero if the given site itself has nonzero initial occupation) in the main govern the nonlinear dynamics of the system, whereas the $U/J$ ratio governs the global structure of motion (localized vs. superfluid-like). It remains to understand why the resonance seems unimportant for the development of chaos, and what -- if any -- influence on dynamics is exerted by the Mott/fluid transition (the Lyapunov exponents show no sign of abrupt change for any $U/J$ value).

%Some sites (1, 3 and 9 in Fig.~\ref{figtabtablesingle} and 9 in Fig.~\ref{figtabtabledouble}) show "lines of weak chaos" in the parameter space: trajectories $U(\mu)$ where chaos remains weak even though slightly off the line it can be very strong. We postpone the research of this phenomenon for further work.

%Running a bit ahead (to the next section) we will see that the distance from the nearest filled site is also crucial for chaotic diffusion; the transition to \emph{strong chaos at long times} is however also determined by the resonances \footnote{Indeed we have already seen that resonance-protected orbits require longer times to become fully chaotic; it turns out that once this happens the chaos is always stronger in the resonant case.}.

Our results on the Lyapunov exponents show general agreement with the classical and quantum indicators of chaos studied in \cite{Nakerst:2022prc,PauschOptimalRoute,PauschSpectrum}: chaos is strongest for intermediate $U/J$ values, and there are "optimal" and "non-optimal" routes to chaos, seen as bright orange and blue lines in Figs.~\ref{figtabtablesingle} and \ref{figtabtabledouble}.  However, the strong dependence on the initial conditions was not discussed so far to the best of our knowledge. It can only happen in systems with mixed phase space \footnote{If there is a single connected chaotic component encompassing most of the phase space, one expects that every orbits will eventually explore most of the phase space so the dynamics will not strongly depend on the initial conditions.}, hence the phase space of the Bose-Hubbard Hamiltonian (at least in the classical limit) remains mixed even in the true many-body regime (we mainly show $L=10$ results for convenience but in the Appendix \ref{secappmore} we show that the same behavior occurs also for chains with up to $100$ sites).

Another property obvious from Figs.~\ref{figtable}-\ref{figtabtabledouble} is that generically many (indeed, almost $L$) Lyapunov exponents are positive \footnote{of course, since the system is Hamiltonian, for every positive exponent there will be another with the opposite sign; together with the energy conservation, this means that the total number of positive exponents is at most $L-1$.}. In other words, the system is hyperchaotic, which was already found for the Bose-Hubbard model (though with slightly different interactions) in \cite{McCormack:2021Photo...8..554M}. The strong hyperchaos that we see is in accordance with the situation for localized states in \cite{McCormack:2021Photo...8..554M}; we did not try to reproduce these exact solutions (and the interaction terms differ anyway) but we suspect that generically, in a large ($L\gg 1$) Bose-Hubbard system, most of the exponents will be positive unless we specially choose a solution corresponding to the minimum of energy.

Finally, one is often interested in the \emph{maximum} Lyapunov exponent rather than the exponents for individual variables. For our purpuses it is more informative to look at the exponents corresponding to individual actions $I_n$ but for completeness we give also the maximum exponent in Appendix \ref{secappmaxle}.

\section{\label{secdiff}Diffusion: anomalous and normal}

\subsection{Anomalous diffusion}

In this section we study the evolution of distribution functions in phase space $\mathcal{P}(I_n,\phi_n;t)$, and in particular the broadening of the distribution due to diffusion. This will lead us to the central phenomenon of interest for this paper -- the anomalous scaling laws for the second moment of the distribution.
More specifically, we consider a population of orbits with initial conditions drawn from a distribution which is initially sharply peaked around some point $(I_n^{(0)},\phi_n^{(0)})$ \footnote{We have tried uniform and Gaussian distributions and the outcome is the same. The dependence on the initial value of the variance is likewise very weak, as long as it is at least an order of magnitude smaller than unity.}, and look at the time evolution of the second moment (variance) of the action: 
\begin{equation}
\sigma^2(I_n)\equiv\langle I_n^2(t)\rangle-\langle I_n(t)\rangle^2,
\end{equation}
where the left-hand side $\sigma^2(I_n)$ is obviously also time-dependent (although we do not write explicitly $\sigma^2(I_n)(t)$ in order not to crowd the notation). To remind, normal diffusion, driven by an uncorrelated random-walk process (and described by the Langevin equation with white noise \cite{CohenElliott:2015book} or alternatively by the (normal) Fokker-Planck equation \cite{Risken:1989book}) corresponds to linear growth of the variance: $\sigma^2(I_n)\propto Dt$, where $D$ is the diffusion coefficient. Although typical applications in fluids, plasmas and many-body systems exhibit diffusion in real space, it also happens \emph{in the space of action variables} in the phase space of strongly chaotic Hamiltonian systems \cite{Lichtenberg:1989,ZASLAVSKY2002461}. The rationale is that the evolution of angles for strongly chaotic orbits can be well approximated by an uncorrelated random process, leading to statistically independent increments of the actions at each "time step" (i.e. some typical timescale). Weak chaos and mixed phase space are known to lead to anomalous diffusion, with the scaling $\sigma^2(I_n)\propto t^\zeta$, with $\zeta$ being some arbitrary exponent ($\zeta<1$ corresponding to subdiffusion and $\zeta>1$ tu superdiffusion, including $\zeta=2$ for ballistic flights).

In this subsection we present the numerical results for superdiffusion, i.e. for the exponent $\zeta>1$ observed in our model. A fully general formula for $\zeta$ depending on the parameters and initial conditions is very hard to find but there are general trends:
\begin{enumerate}
\item{The anomalous transport exponents for the action $I_n$ take the values $\zeta_n=2m$ or $\zeta_n=4m$ (for $m=0,1,2,\ldots$) -- they are strictly even integers and come in arithmetic series with spacing $2$ or $4$. We find this very striking -- such integer and strictly "quantized" values are very rare in the literature \cite{Lichtenberg:1989}.}
\item{The value of the exponent $\zeta_n$ for each site depends on the initial conditions, that is on the distribution of filled and empty sites. If the site $n$ is $m$ sites away from the nearest filled site $n_0$ (so that $m=\vert n-n_0\vert$), then the exponent is generically $\zeta_n=4m$.}
\item{In particular, the filled site has $m=0$ -- the distance from the nearest filled site is zero, thus $\zeta_n=0$ -- the distribution function spreads at most logarithmically.}
\item{If partially filled sites with similar occupation numbers are present (so that the notion of distance to the nearest filled site is vague), then the exponents take the form $2m$.}
\item{The endpoints of the chain act effectively as filled sites, meaning that the relation $m=\vert n-n_0\vert$ is modified as $m=\mathrm{min}\left(\vert n-n_0\vert,n,L-n\right)$.}
\end{enumerate}
From the above, each site $n$ has in principle its own exponent $\zeta_n$ but of course some degeneracy will be present, as for some $n,n'$ one may have $\zeta_n=\zeta_{n'}$. Thus in general we do not obtain as many scalings as there are actions. Once again, remember that transport in the action space (the typical subject of statistical studies of chaos \cite{Lichtenberg:1989}) in the Bose-Hubbard model simply describes the distribution of the occupation numbers along the chain. 

\begin{figure}[H] 
\centering
\includegraphics[width=.9\linewidth]{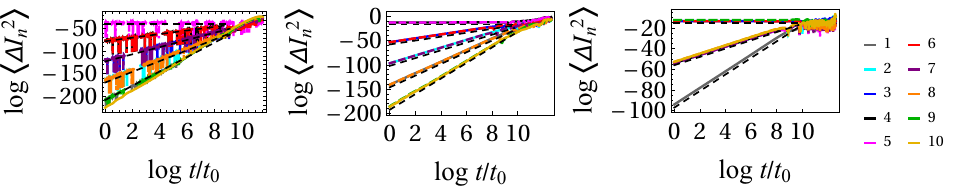}
\caption{Log-log plot of the second central moment (variance) of the actions for an ensemble of orbits in the chain of length $L=10$, with $U/J=25,\mu/J=0.05$. Initially the filled sites are $n=5$ (left), $n=4,5$ (center) and $n=3,6,9$ (right). The exponents take values $0,4,8,\ldots$, determined by the distance from the nearest initially occupied site. Black dashed lines are analytic plots $\langle\Delta I_n^2\rangle\propto t^{4m}$. The color legend for the site numbers is on the right.}
\label{figanomdiff1}
\end{figure}

\begin{figure}[H] 
\centering
\includegraphics[width=.9\linewidth]{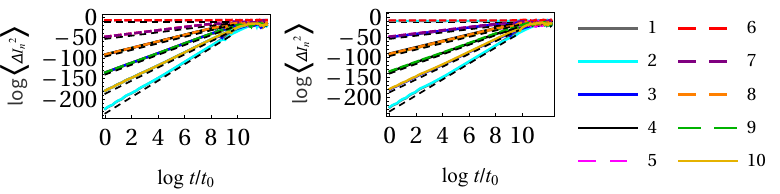}
\caption{Log-log plot of the second central moment (variance) of the actions for an ensemble of orbits in the chain of length $L=20$, with $U/J=25,\mu/J=0.05$. Initially filled sites are $n=7,16$ (left) and $n=7,12,16$ (right). The number of exponents is larger for longer chains. The color legend for the site numbers is on the right.}
\label{figanomdiff2}
\end{figure}

\begin{figure}[H] 
\centering
\includegraphics[width=.9\linewidth]{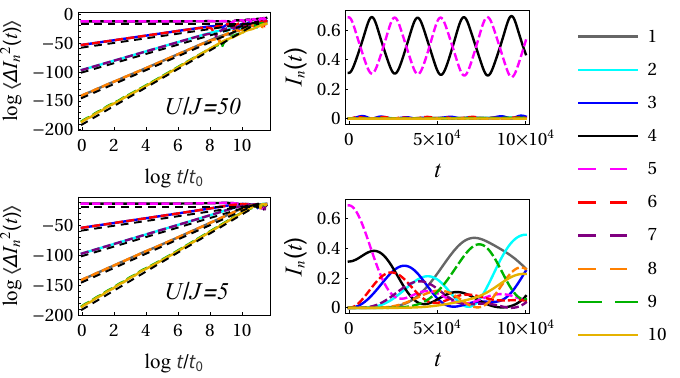}
\caption{Log-log plot of the variance of the actions for an ensemble of orbits (left) and an example of a single orbit (right) in the chain of length $L=10$, with $U/J=50$ (top) and $U/J=5$ (bottom), with $\mu=0$. Initially filled sites are $n=4,5$. Although the top row shows Mott-like localization and the bottom row the superfluid regime, both regimes exhibit the same scaling laws for anomalous diffusion. The color legend for the site numbers is on the right.}
\label{figanomdiffu}
\end{figure}

Examples of the above findings are found in Fig.~\ref{figanomdiff1}. For the initial conditions with one, two and three filled sites in a chain with $L=10$ wells we see the $4m$ scaling exponents, and also their dependence on the relative positions of filled sites. In Fig.~\ref{figanomdiff2} we further corroborate our claims by finding the same behavior for longer chains ($L=20$). In Fig.~\ref{figanomdiffu} we specifically show that the same scaling behavior is found independently of the ratio $U/J$, i.e. independently of the Mott-like/fluid-like regime and the degree of nonintegrability \footnote{Remember that for $U/J$ very small or very large the system becomes integrable.}. All figures are given as log-log plots in order to clearly show the power-law scaling.

Now we show the cases where the series $2m$ appears. This happens for sufficiently complicated initial conditions, when the chain abounds with partially filled and near-resonant sites. Fig.~\ref{figanomdiff2m} brings a few examples; more examples can be found in Appendix \ref{secappmore}.

%In Figs.~\ref{figanomdiff1} and \ref{figanomdiffu} the sites are initially all in $1:1$ resonances, i.e. with equal actions (except the single-site case where this is obviously impossible). In Fig.~\ref{figanomdiff2} the initial conditions are nonresonant. In this case the resonance does not influence the scaling laws, it may only reduce the number of exponents (because the notion of the "nearest site with higher occupation number" is nonunique if there are sites with equal occupation).

\begin{figure}[H] 
\centering
\includegraphics[width=.9\linewidth]{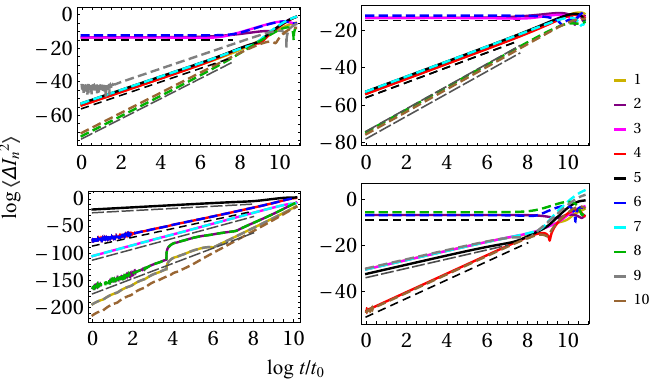}
\caption{Log-log plot of the second central moment (variance) of the actions for an ensemble of orbits in the chain of length $L=10$ at $\mu/J=0.25$ and $U/J=1$ (left) vs. $U/J=5$ (right), with four/six partially occupied sites at $t=0$ (top/bottom row). The scaling exponents are now a mix of $4m$ and $4m+2$ values, outlined by black and gray dashed lines respectively. The additional exponents are obtained in the presence of multiple partially filled sites with similar actions (occupation numbers). The color legend for the site numbers is on the right.}
\label{figanomdiff2m}
\end{figure}

\begin{figure}[H] 
\centering
\includegraphics[width=.9\linewidth]{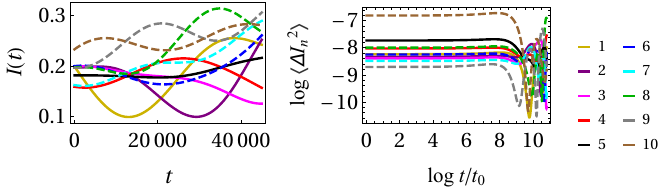}
\caption{Evolution of the actions $I_n(t)$ (left) and of the second central moment (variance) of the actions for an ensemble of orbits $\langle I_n^2(t)\rangle$ (right) for the chain of length $L=20$, with $U/J=5$, $\mu/J=0.25$ for the homogeneous initial distribution (the total occupation is equally distributed among all sites). In this case there is no transport at all, in accordance with the claim that the scaling exponents are determined by the distance to the nearest filled site -- now all sites have equal fillings. The color legend for the site numbers is on the right.}
\label{fignodiff}
\end{figure}

In general, the number of exponents $\zeta_n$ decreases as the percentage of occupied sites grows. In particular, for a completely homogeneous chain there is no transport at all (Fig.~\ref{fignodiff}). This can be explained in the following way. Given the number of initially occupied sites $k$ one can look at the $k+1$ segments of the chain defined by the occupied sites. This means that our system can be effectively divided into a series of smaller mutually connected chains. Sites symmetric to each other with respect to the initially occupied site behave in the same way. This is why the edge effects are so important, as they dictate the transport through the subchains.

Finally, we note that for periodic boundary conditions we observe no transport at all (for any initial condition and for all parameter values). The explanation is easy: we have seen that the sites symmetric with respect to the initially occupied one behave in the same way (for example, if the site $n$ was initially occupied, it gives rives to same exponent $4n$ for the sites $n-1$ and $n+1$). In the ring case this logic does not hold as the distances between the sites are non-unique, the system equilibrates very quickly and does not undergo anomalous diffusion.

\subsubsection{Scaling analysis and the anomalous transport exponents}

It is notoriously difficult to obtain the anomalous diffusion coefficients analytically \cite{METZLER20001,zaslavsky2007physics}: they are a consequence of strong long-range and long-time correlations, resulting in integro-differential equations for the kinetics (which are hard even to write down, let alone solve). We can only offer a very crude scaling analysis based on the general considerations of \cite{Zaslavsky1994FractionalKE}.

The transport in action space around some initial values $\mathbf{I}(t=0)$ is dictated by the leading angle-dependent term in the perturbative expansion around $\mathbf{I}$ (the angle-independent terms describe integrable dynamics and cannot change the actions). Away from resonances, the leading transport-inducing terms are contained in $H_2$ in Eq.~(\ref{hpert}), with the coefficients given by Eqs.~(\ref{h2}-\ref{h2tilde}). If we start from a nonzero value for some $I_j$ with $I_{j\pm 1}$ being initially much smaller or zero, we may regard $I_{j\pm 1}$ as the only dynamical degree of freedom at early times while $I_j\sim\mathrm{const.}$ (because the relative change of $I_j$ is initially negligible). The total Hamiltonian $H'$ in Eq.~(\ref{hpert}) then has the form of a pendulum for $I_{j+1}$ or $I_{j-1}$, with mass term proportional to $U/2$ and the potential term behaving as $-2J^2/(UI_j)$ (at leading order in $I_j$). The characteristic timescale is given by the period of the pendulum \footnote{Of course, the pendulum approximation is not exact and the motion is not periodic, but as a crude timescale estimate this logic should work.}:
\begin{equation}
T\sim\frac{\sqrt{I_j}}{J}.
\end{equation}
Now if we rescale $(T,I_j)\mapsto (T\lambda_T,I_j\lambda_\ell)$, the above equation will be invariant if
\begin{equation}
\lambda_\ell=\lambda_T^2.
\end{equation}
This is the scaling law for transport from $I_j$ to an initially (near-)empty neighboring site. If we look at the transport from $j$-th to $j\pm m$-th site instead, we have the same scaling for every site, but we now have $m$ actions involved in transport (from $I_j$ to $I_{j+1}$, from $I_{j+1}$ to $I_{j+2}$, etc.) so in the $m$-dimensional action subspace we have $\lambda_I=\lambda_\ell^m$, where $\lambda_I$ is the scaling exponent for the action transport (if $m=1$ then $\lambda_I=\lambda_\ell$ but otherwise they are not equal because the trajectory exists in some higher-dimensional space). This implies
\begin{equation}
\lambda_I=\lambda_\ell^m=\lambda_T^{2m}.
\end{equation}
From the Renormalization Group of Kinetics formalism in \cite{Zaslavsky1994FractionalKE,ZASLAVSKY2002461,zaslavsky2007physics}, the anomalous diffusion exponent is
\begin{equation}
\zeta_m=\frac{2\log\lambda_I}{\log\lambda_T}=4m,
\end{equation}
in accordance with the numerical results.

The perturbative Hamiltonian considered above does not make sense near a resonance even though even in resonant cases we usually still see the $4m$ series, except if many resonances are present. We do not have a good understanding of this. For the resonant case, the resonant Hamiltonian in Eq.~(\ref{hpertres}) has no discrete scaling symmetry: assuming that $I_0\gg I_r$, the leading term scales as $I_0$ and the subleading (potential) term as $\sqrt{I_0}$. Hence the only scaling is $(T,I_0)\mapsto (T\lambda,I_0\lambda)$, for characteristic time and action scales $T$ and $I_0$ and some rescaling factor $\lambda$; hence $\lambda_T=\lambda_\ell$. For the actions, we have $\lambda_I\sim\lambda_\ell^m$ just as in the previous case, therefore
\begin{equation}
\zeta_m=\frac{2\log\lambda_I}{\log\lambda_T}=\frac{2\log\lambda^m}{\log\lambda}=2m.\label{scalingres}
\end{equation}
This is the same exponent as in the numerics, however, as we said above, it does not happen whenever there is a resonance (usually the resonances also have $4m$) but only for some complicated and rather fine-tuned configurations. We hope to gain a better understanding of these phenomena in future work.
 
\subsection{\label{secdifflang}Normal diffusion and Langevin equation}

\begin{figure}[H] 
\centering
\includegraphics[width=.9\linewidth]{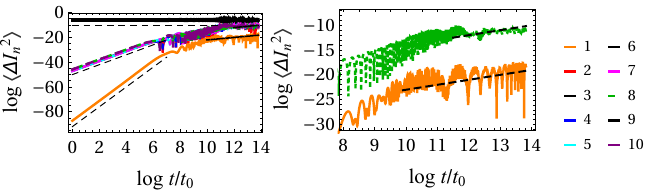}
\caption{Evolution of the variance of the actions for an ensemble of orbits $\langle I_n^2(t)\rangle$ for the chain of length $L=10$, with $U/J=10$, $\mu/J=0$. At earlier times ($t/t_0\lesssim\exp(7)$) there is the usual robust anomalous diffusion, with exponents $\zeta=0,4,8$ (dashed lines in the left panel). After a fluctuating transitional period $\exp(7)\lesssim t/t_0\lesssim\exp(10)$, the initially non-filled sites exhibit normal diffusion with exponent $\zeta_n=1$, i.e. $\langle I_n^2(t)\propto t\rangle$ (dotted lines in the left panel and the zoom-in of the late-time epoch for $n=1,8$ in the right panel).}
\label{fignormdiff}
\end{figure}

The findings of superdiffusion in the previous subsection hold only up to some epoch $t_*$ which in fact can be quite large, of the order $10^5\times t_0$. For $t>t_*$, a short transitional regime with no clear trends in the time evolution of the variance is followed by the epoch of \emph{normal diffusion} with $\zeta_n=1$ starting around some $t=t_{**}$ where $t_{**}$ can be very long \footnote{Indeed, in many integrations the numerics becomes unstable before reaching $t_{**}$; but we expect that normal diffusion will always exist once the ensemble of orbits has explored sufficiently large volume of the phase space.}. One numerical example is given in Fig.~\ref{fignormdiff}.

Normal chaotic diffusion is quite well-studied starting from the pioneering work of Chirikov \cite{Chirikov}; a standard introduction to the Chirikov theory and its later developments is found in \cite{Lichtenberg:1989}. The Chirikov approach rests on assuming completely uncorrelated dynamics of angle variables and the averaging of the equations of motion for the actions over the angles. While the leading-order result is very simple and transparent, the Chirikov approach becomes more and more complicated if higher-order corrections (i.e. correlations between subsequent "timesteps") are to be included. We will thus only compute the lowest-order result following \cite{Chirikov,Lichtenberg:1989}, and then we will show how to obtain a more systematic approximation scheme from the Langevin equation formalism \cite{Kawasaki:1973,Dengler:2015}.

\subsubsection{Leading-order diffusion matrix}

We have $L$ actions, thus diffusion is described by the diffusion matrix $\hat{D}$ of size $L\times L$. However, the constraint reduces the total number of independent degrees of freedom to $L-1$. Since the constraint can be solved explicitly, it poses no problems: we simply replace $I_L$ by $1-\sum_{j=1}^{L-1}I_j$ and consider the system as a function of $I_1,\ldots I_{L-1}$. We now have the reduced diffusion matrix $\hat{\bar{D}}$ of size $(L-1)\times (L-1)$, with elements $\bar{D}_{ij}$ for $1\leq i,j\leq L-1$. Following \cite{Chirikov} and the basic notions of kinetics, the element $\bar{D}_{ij}$ is given by the average of the right-hand sides of the equations of motion:
\begin{equation}
\bar{D}_{ij}=\langle\langle\dot{I}_i\dot{I}_j\rangle\rangle,
\end{equation}
where $\langle\langle (\ldots)\rangle\rangle$ denotes the averaging over the angle variables from $0$ to $2\pi$ \footnote{We deliberately differentiate between the angle average and the average over the ensemble $\langle(\ldots)\rangle$. Although the two have to coincide if the diffusion approximation is \emph{exact}, it never is in practice.}. It is easy to check that most matrix entries are zero. Explicitly, the elements are
\begin{eqnarray}
\bar{D}_{n,n+1}&=&\frac{1}{(2\pi)^3}\int_0^{2\pi}d\phi_n\int_0^{2\pi}d\phi_{n+1}\int_0^{2\pi}d\phi_{n+2}\dot{I}_n\dot{I}_{n+1}=-\frac{1}{2}\sqrt{I_nI_{n+1}}\nonumber\\
\bar{D}_{n,n}&=&\frac{1}{(2\pi)^3}\int_0^{2\pi}d\phi_{n-1}\int_0^{2\pi}d\phi_n\int_0^{2\pi}d\phi_{n+1}\dot{I}_n^2=\frac{1}{2}\left(I_{n-1}+I_{n+1}\right)\nonumber\\
\bar{D}_{n,n-1}&=&\frac{1}{(2\pi)^3}\int_0^{2\pi}d\phi_n\int_0^{2\pi}d\phi_{n-1}\int_0^{2\pi}d\phi_{n-2}\dot{I}_{n-1}\dot{I}_n=-\frac{1}{2}\sqrt{I_{n-1}I_n}\nonumber\\
\bar{D}_{i,j}&=&0,~~~\vert i-j\vert>1.\label{diffcoeff0}
\end{eqnarray}
The diffusion coefficient matrix is only nonzero on, right above and right below the main diagonal. This is in line with the locality of interactions -- the actions are interpreted as occupation numbers and particles have to travel from one site to its neighbors, not jump to a non-neighboring site. 

A complete solution of the resulting multidimensional diffusion equation with hard-wall boundary conditions is quite complicated (although there are no problems of principle in finding it). However, in this paper we are mainly interested in the diffusion coefficients themselves, since the rate of growth of the variances $\langle\Delta I_n^2\rangle$ is determined solely by the diffusion coefficients. Before we compare the numerical data with the theoretical prediction, we will construct also the Langevin equation which provides systematic corrections to (\ref{diffcoeff0}).

\subsubsection{Langevin equation}

Assuming that the evolution of angles (fast variables) is well-described by a Wiener process, the resulting equations of motion for the actions (slow variables) acquire the form of Langevin equation. This is another well-known way to obtain the diffusion equation \cite{Kawasaki:1973} and it allows us to go beyond leading-order estimate for the diffusion matrix as we can systematically include the correlations between angles at different times. Clearly, the Langevin approach also works only for orbits that spend most time in the chaotic sea; the phase space is generically mixed as we have shown numerically (Figs.~\ref{figorbits},\ref{figorbitsres}) and many orbits do not obey the Langevin kinetics or only obey it after long transient times \footnote{After all, this is obvious already from the fact that many actions exhibit strongly anomalous diffusion, at least until some late time.}.

We work in the Ito formalism. The starting point are the equations of motion (\ref{eomi}-\ref{eomphi}), with the constraint (\ref{constrainti}). In order to simplify the calculations and get rid of the square roots of actions, we introduce new action variables (still with the constraint):
\begin{equation}
a_j\equiv\sqrt{I_j},~~\sum_{j=1}^La_j^2=1,~~~~j=1,\ldots L.
\end{equation}
It is easy to rewrite Eqs.~(\ref{eomi}-\ref{eomphi}) in terms of $(a_j,\phi_j)$:
\begin{eqnarray}
\label{eomaphi}\dot{\phi_j}&=&-\mu+U a_j^2-J\left(\frac{a_{j-1}}{a_j}\cos\left(\phi_j-\phi_{j-1}\right)+\frac{a_{j+1}}{a_j}\cos\left(\phi_j-\phi_{j+1}\right)\right)\\
\label{eomaa}\dot{a_j}&=&J\left[a_{j-1}\sin\left(\phi_{j-1}-\phi_j\right)+a_{j+1}\sin\left(\phi_{j+1}-\phi_j\right)\right].
\end{eqnarray}
We now assume that the angles $\phi_j$ can be approximated by Gaussian random noise \footnote{We cannot provide a proof of this assumption (which in any case certainly only holds approximately); a look at the strongly erratic oscillations of some orbits in Figs.~\ref{figorbits} and \ref{figorbitsres} provides some justification.}, which means that sines of their difference also have finite second moment and define a Wiener process. Eq.~(\ref{eomaphi}) is thus dropped in the Langevin approach (we do not solve for $\phi_j$ but approximate it by noise), and in the second row the sines of the angle differences become increments $dW_j$ of a Wiener process, yielding the equation (it is again convenient to write the variables as matrices with two indices $i,j=1,\ldots L$, denoted also by hatted letters):
%\begin{equation}
%\label{langa}da_i=g_{ij}dW_j,~~\hat{g}=J\left(\begin{matrix}a_2 & \ldots & \ldots & -a_L\\ -a_1 & a_3 & \ldots & \ldots\\ 0 & -a_2 & a_4 & \ldots\\ \ldots & \ldots & \ldots & \ldots\\ \ldots & -a_{L-2} & a_L & 0\\ \ldots & \ldots & -a_{L-1} & a_1 \end{matrix}\right)_{L\times L},~~~\sum_{j=1}^La_j^2=1.
%\end{equation}
\begin{equation}
\label{langa}da_i=g_{ij}dW_j,~~\hat{g}=J\left(\begin{matrix}0 & a_2 & 0 & \ldots & \ldots & \ldots\\ -a_1 & 0 & a_3 & 0 & \ldots & \ldots\\ 0 & -a_2 & 0 & a_4 & 0 & \ldots\\ \ldots & \ldots & \ldots & \ldots & \ldots & \ldots \\ \ldots & \ldots & 0 & -a_{L-2} & 0 & a_L\\ \ldots & \ldots & \ldots & 0 & -a_{L-1} & 0 \end{matrix}\right)_{L\times L},~~~\sum_{j=1}^La_j^2=1.
\end{equation}
Notice that there is no purely deterministic term (the whole right-hand side is proportional to $dW_j$) and that the Langevin system is still constrained by the overall normalization constraint. The easiest way to proceed is again to solve the constraint and express $a_L=\sqrt{1-\sum_{i=1}^{L-1}a_i^2}$ in $g_{ij}$. In this way we eliminate $(a_L,\phi_L)$, reducing the matrix of the system from size $L\times L$ to size $(L-1)\times (L-1)$, with the variables
\begin{equation}
\bar{a}_i\equiv(a_1,\ldots a_{L-1}),~~d\bar{W}_i\equiv(dW_1,\ldots dW_{L-1}),\label{atilde}
\end{equation}
which satisfy the equation
\begin{equation}
%\nonumber d\bar{a}_i=\bar{g}_{ij}d\bar{W}_j,~~\hat{\bar{g}}=J\left(\begin{matrix}a_2 & \ldots & \ldots\\ -a_1 & a_3 & \ldots & \ldots\\ 0 & -a_2 & a_4 & \ldots\\ \ldots & \ldots & \ldots & \ldots\\ \ldots & \ldots & -a_{L-2} & \sqrt{1-\sum_{i=1}^{L-1}a_i^2}\end{matrix}\right)_{(L-1)\times (L-1)}.
d\bar{a}_i=\bar{g}_{ij}d\bar{W}_j,~~\hat{\bar{g}}=J\left(\begin{matrix}0 & a_2 & 0 & \ldots & \ldots & \ldots\\ -a_1 & 0 & a_3 & 0 & \ldots & \ldots\\ 0 & -a_2 & 0 & a_4 & 0 & \ldots\\ \ldots & \ldots & \ldots & \ldots & \ldots & \ldots \\ \ldots & \ldots & 0 & -a_{L-3} & 0 & a_{L-1} \\ \ldots & \ldots & \ldots & 0 & -a_{L-2} & 0\end{matrix}\right)_{(L-1)\times (L-1)}\label{langatilde}
\end{equation}
Ideally, we would like to transform (\ref{langatilde}) to the "canonical" Langevin form, where the right-hand side is $\mathbf{\bar{a}}$-independent and in general equal to the sum of a deterministic term and a noise term \cite{CohenElliott:2015book}. In Appendix \ref{secapplang} we show that for the above equation this is not possible. This in turn means that (\ref{langatilde}) is not \emph{exactly} equivalent to a diffusion equation; even under the assumption of uncorrelated angles there is no purely diffusive behavior.
 
However, we can still integrate Eq.~(\ref{langatilde}) perturbatively, order by order. It will turn out that the lowest nonzero term is the quadratic term, corresponding precisely to normal diffusion. Expanding the equation around $\bar{a}_i^{(0)}\equiv\bar{a}_i(0)$ and likewise $\bar{g}^{(0)}_{ij}\equiv\bar{g}_{ij}\left(\bar{a}\left(0\right)\right)$ we get:
\begin{eqnarray}
\bar{a}_i(t)&=&\bar{a}_i^{(0)}+\int_0^t\bar{g}_{ij}\left(\bar{a}\left(t'\right)\right)d\bar{W}_j(t')=\nonumber\\
&=&\bar{a}_i^{(0)}+\bar{g}_{ij}^{(0)}\int_0^td\bar{W}_j(t')+\frac{\partial g^{(0)}_{ij}}{\partial\bar{a}_k}\int_0^t\left(\bar{a}_k\left(t'\right)-\bar{a}_k\left(0\right)\right)d\bar{W}_j(t')=\nonumber\\
&=&\bar{a}_i^{(0)}+\bar{g}_{ij}^{(0)}\int_0^td\bar{W}_j(t')+\frac{\partial\bar{g}^{(0)}_{ij}}{\partial\bar{a}_k}\bar{g}_{kl}^{(0)}\int_0^{t'}\int_0^td\bar{W}_l(t'')d\bar{W}_j(t').\label{langasol}
\end{eqnarray}
Assuming that the Wiener process is symmetric and white-in-time so that
\begin{equation}
\langle W_i(t)\rangle=0,~~\langle W_i(t)W_j(t')\rangle=\delta(t-t')\sigma^2_{ij},
\end{equation}
we can calculate the expectation values of the first and second moment of $\bar{a}$:
\begin{eqnarray}
\langle\bar{a}_n\left(t\right)-\bar{a}_n\left(0\right)\rangle&=&0\label{langmom1}\\
\langle\left(\bar{a}_n\left(t\right)-\bar{a}_n\left(0\right)\right)\left(\bar{a}_m\left(t\right)-\bar{a}_m\left(0\right)\right)\rangle&=&\frac{1}{2}\bar{g}^{(0)}_{ni}\bar{g}^{(0)}_{mj}\sigma^2_{ij}t+O\left(\sigma^4\right).\label{langmom2}
\end{eqnarray}
Here, $\sigma_{ij}^2$ are the variances of the Wiener process and by $\sigma^4$ we have schematically denoted the fourth-order momenta of the process (which are not necessarily all equal). From (\ref{langmom2}), we get the diffusion coefficient matrix:
\begin{equation}
\bar{D}_{nm}=\bar{g}^{(0)}_{ni}\bar{g}^{(0)}_{mj}\sigma^2_{ij}.\label{diffcoeff}
\end{equation}
The only remaining step is to compute the second moments of the Wiener process, which are obtained by averaging and normalizing \footnote{The normalization by the factor $\langle\langle\dot{\phi}_i^2\rangle\rangle$ is necessary because by construction of the Langevin equation (\ref{langa}) the Wiener process is assumed to be normalized, i.e. it is already multiplied by the right-hand side of the equations of motion.} the right-hand side of the expressions for $\dot{\phi}_i\dot{\phi}_j$ in (\ref{eomaphi}) over all the angles. However, unlike equal-time averages of $\dot{I}_i\dot{I}_j$ that yield the leading-order expression (\ref{diffcoeff0}), which become zero if $\vert i-j\vert>1$, here even at equal times all averages are nonzero: the Wiener process is white-in-time but has long-range correlations.

\begin{figure}[H] 
\centering
\includegraphics[width=.45\linewidth]{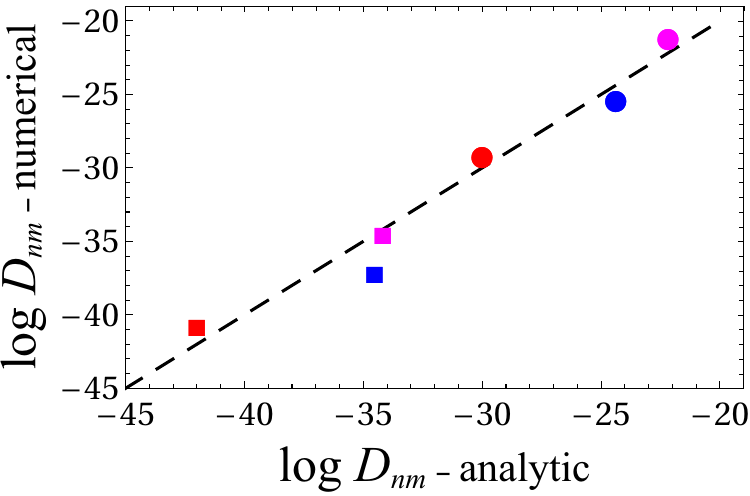}
\caption{Logarithms of diffusion coefficients between the sites $n$ and $m=n+1$ (circles) and between the sites $n$ and $m=n+2$ (squares) obtained by fitting numerically the growth rate of the action variances $\langle\Delta I_n^2\rangle$, for configurations with $U/J=50,10,1$ (blue, magenta, red respectively) and $\mu/J=0$ in a chain of length $L=20$ with initially occupied sites $n=4$ and $n=5$, plotted against the predictions of the next-to-leading order estimate (\ref{difflang1}). The black dashed curve corresponds to ideal match (analytic=numerical). The agreement is far from perfect (bear in mind the scale is logarithmic) but is in fact unexpectedly good for such a simple model.}
\label{figdiffcoeff}
\end{figure}

The local contribution to the diffusion matrix is
\begin{equation}
%\sigma_{ii}^2\equiv\langle\langle\dot{\phi}_i^2\rangle\rangle=\frac{1}{(2\pi)^3}\int_0^{2\pi}d\phi_i\int_0^{2\pi}d\phi_{i-1}\int_0^{2\pi}d\phi_{i+1}\dot{\phi}_i^2=\frac{J^2}{2}\left(I_{j-1}+I_{j+1}\right),\label{difflang0}
\sigma_{ii}^2\equiv\frac{\langle\langle\dot{\phi}_i^2\rangle\rangle}{\langle\langle\dot{\phi}_i^2\rangle\rangle}=1,\label{difflang0}
\end{equation}
and it yields exactly the same result as (\ref{diffcoeff0}) upon plugging in into (\ref{diffcoeff}). This is expected -- the local average of the angle already captures the nearest-neighbor transport of the actions. The next-order corrections read (we change back from the computationally convenient $a_j$ variables to the more natural $I_j$ variables):
\begin{equation}
%\sigma_{i,i+1}^2&=&\equiv\frac{\langle\langle\dot{\phi}_i\dot{\phi}_{i+1}\rangle\rangle}{\langle\langle\dot{\phi}_i^2\rangle\rangle}=UI_i\frac{2\mu-UI_i}{J^2+\left(\mu-UI_i\right)^2}\nonumber\\
\sigma_{i,i\pm 1}^2\equiv\frac{\langle\langle\dot{\phi}_i\dot{\phi}_{i\pm 1}\rangle\rangle}{\langle\langle\dot{\phi}_i^2\rangle\rangle}=UI_i\frac{2\mu-UI_i}{J^2+\left(\mu-UI_i\right)^2}\label{difflang1}.
\end{equation}
The leading terms are now linear in the actions. Higher-order corrections would involve higher-order terms in $I_i$ for various $i$, however it does not make sense to go to arbitrarily high orders as the Langevin approximation is itself of limited validity. We now compare the numerical fits of the diffusion coefficient to the analytic estimates in Fig.~\ref{figdiffcoeff}. The next-to-leading order approximation (\ref{difflang1}) yields at least a good order-of-magnitude estimate -- perhaps not an impressive result by itself but we find it encouraging bearing in mind the simplicity of the analytical model.

Another way to understand transport, in particular the diffusive regime, is to start from DNSE which is analytically solvable for $J$ small. We have not made much progress in this direction but some thoughts can be found in Appendix \ref{secappnlse}.

\section{\label{secconc}Discussion and conclusions}

Our key result is the existence of two regimes in the spread of an initial ensemble of orbits: at early times there is anomalous diffusion with a strikingly regular series of exponents, equal to $4m$ or $2m$ ($m=1,2,\ldots$) depending on the initial conditions, which at late times crosses over to normal diffusion with transport exponent $1$. This might correspond to the thermalization of the system, which we plan to check by computing the thermodynamic potentials. Also, it is intriguing to know what this superdiffusion-to-normal-diffusion crossover means for the quantum dynamics.

One might be disappointed that the results we have so far are purely classical: the really exciting and experimentally relevant physics is in the quantum regime. Nevertheless, the classical regime is often the key to a controlled treatment of quantum dynamics, as the latter is very hard to study \emph{ab initio} and without approximations. In \cite{PolkovnikSachdev2002,Polkovnik2003Main,Graefe:2007,Trimborn:2008,Nakerst:2022prc}, the authors have arrived at important conclusions for the quantum model from the classical equations. In particular, \cite{Graefe:2007,Trimborn:2008} have established the connection between the classical regime and the Wigner and Husimi functions which provide important insight into quantum decoherence and quantum chaos. We plan to move in this direction in further work. Within the formalism of the truncated Wigner approximation, classical distribution functions that we study in detail in this work provide the initial condition for the evolution of the Wigner function. In further work we will try to understand the consequences of classical superdiffusion for the evolution of Wigner function and decoherence, an important topic of interest in recent works \cite{Dahan:2022classical,Haque:PhysRevE.89.042112,Polkovnik2018,Sorg:PhysRevA.90.033606,rizzatti2020double}.

It is unclear how far one can go in analytical work on this topic. We have provided a very simple derivation of the superdiffusion coefficients but it is just a back-of-the-envelope calculation which likely does not explain everything, in particular the transition from the $4m$ series to the $2m$ series of exponents. On the other hand, the normal diffusion regime is well-described by the usual Langevin formalism; while high quantitative accuracy is not easy to achieve, the essential physics is certainly captured. 

Finally, our results bring some unexpected conclusions about the nature of chaos in many-body systems. Despite the common expectation that systems with few degrees of freedom exhibit mixed (regular/chaotic) phase space while the many-degrees-of-freedom systems (when they are nonintegrable) show developed chaos because of many interacting modes, this is not so in the Bose-Hubbard model: there are always weakly chaotic orbits, and the strength of chaos is mainly determined by occupation numbers, not by nonintegrability parameter $U/J$ (where both $U/J\to 0$ and $U/J\to\infty$ correspond to integrable cases and the regime $U\sim J$ is expected to be "most nonintegrable") or the transition between the Mott and the superfluid regime (the transition between the two is obvious when looking at \emph{shapes of single orbits}, but not in their chaoticity or in the spread of the ensemble). One might think that additional conservation laws effectively reduce the number of degrees of freedom in our equations, but the only conservation laws are the conservation of energy (as in any Hamiltonian system) and the number conservation, leading to the constraint (Eqs.~\ref{constraint} and \ref{constrainti}) that we explicitly take into account. These two constraints alltogether reduce the dimension of phase space from $2L$ to $2L-2$ which is by itself insufficient to explain the persistence of anomalous transport even for $L=10$ or even $L=100$. There might be \emph{pseudointegrals of motion}, leading to approximate conservation laws and approximately confining the dynamics to a low-dimensional manifold in phase space. But these are notoriously hard to find, thus for now we do not know if this is the explanation; in the literature, to the best of our knowledge, nothing is known about such pseudointegrals for the Bose-Hubbard system.

All of this is quite surprising and we hope to provide a quantitative explanation in further work.

\acknowledgments

We are grateful to Marco Schiro, Jak\v{s}a Vu\v{c}i\v{c}evi\'c, Fabrizio Minganti and Filippo Ferrari for stimulating discussions. Work at the Institute of Physics is funded by the Ministry of Education, Science and Technological Development and by the Science Fund of the Republic of Serbia. M.~\v{C}. would like to acknowledge the Mainz Institute for Theoretical Physics (MITP) of the Cluster of Excellence PRISMA+ (Project ID 39083149) for hospitality and partial support during the completion of this work.

\appendix

\section{\label{secappdynnum}Numerical methods}

We briefly comment on the numerical procedure used for the calculation of orbits (and also for obtaining the statistical distribution of orbits). The problem reduces to the integration of the equations of motion for the $(P_j,Q_j)$ variables as given in Eq.~(\ref{eompq}). We perform the integrations in the \textrm{Mathematica} package with the routine \textrm{NDSolve}. For each set of integrations one needs to adjust the accuracy and precision goals so that the result becomes independent of their exact values. While it is possible to integrate the equations as a constrained system with the explicit implementation of the constraint (\ref{constraint}), we have found that for most initial conditions, \emph{unconstrained} integration yields results which are very close to those from constrained integration: the violation of the constraint and the relative discrepancies in the values of the variables in the unconstrained integration are both of order of one percent. We have thus often used unconstrained integration, monitoring the value of the constraint; when the difference grows larger than some tolerance we stop the integration. This still allows us to integrate for quite long times, often up to $t=5\times 10^5$ in units of $1/J$, and is much faster than constrained integration. Of course, for every configuration we also perform a few tests with the full constrained integration as a benchmark and sanity check. When the integration is done, we obtain the actions and angles $(I_j,\phi_j)$ from $(P_j,Q_j)$ according to Eq.~\ref{actangle}.

Lyapunov exponents are obtained by writing down and solving the variations of the equations of motion (\ref{eompq}). The equations are linear in the variations $\delta P_j(t),\delta Q_j(t)$, which depend also on the solutions $P_j(t),Q_j(t)$. Therefore, we first solve the equations of motion for $P_j,Q_j$, then we solve also the variational equations \footnote{Here also we do it either with the explicit constraint or as unconstrained integration while monitoring the value of the constraint.}. The results are pretty stable with respect to the initial value of the variation within the interval from $10^{-6}$ to $10^{-14}$, whereas the necessary integration time varies depending on the values of the Lyapunov exponents themselves (but again there is a broad interval of timescales which yield practically the same value of the exponents).

\section{\label{secappmore}Further examples of the scaling laws}

\begin{figure}[H] 
\centering
\includegraphics[width=.9\linewidth]{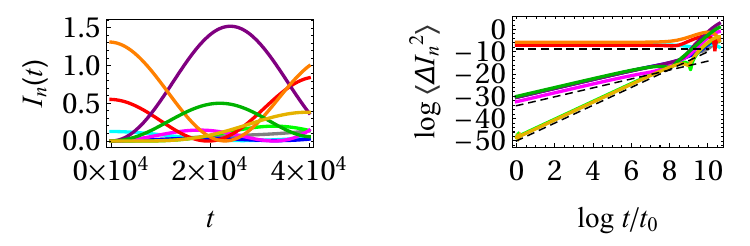}
\caption{Evolution of the actions (left) and the log-log plot of the second central moment (right) of the actions for an ensemble of orbits in the chain of length $L=10$, with $U/J=0.375, \mu/J=0.250$, with initial nonzero actions $I_2,I_6,I_8$. The exponents still obey the general laws we have found. The color code is the same as in Fig.~\ref{fignormdiff}.}
\label{fig3sites}
\end{figure}

\begin{figure}[H] 
\centering
\includegraphics[width=.9\linewidth]{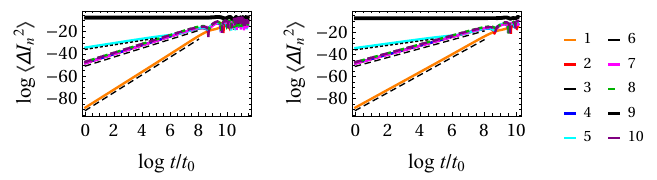}
\caption{Log-log plot of the second central moment (variance) of the actions for an ensemble of orbits in the chain of length $L=10$, with $U/J=50$ (left) and $U/J=5$ (right), with $\mu/J=0.5$. Initially filled sites are $n=3,5,6,9$, with fillings $2/3,0.01,2/3-0.01,2/3$ respectively. Notice how the action $I_5$ starts out with the exponent $\zeta_5=2$ which however later turns into $\zeta_5=4$ as the occupation of this site becomes comparable with the sites $n=3,6,9$ which start which much higher fillings. Unlike the parameters of the system ($\mu,U$), the initial conditions are crucial in determining the anomalous transport.}
\label{figanomdiff2mplus}
\end{figure}

As an illustration of universality, we demonstrate the anomalous diffusion scaling laws for one more distribution of initially filled sites and for a low $U/J$ value, deep in the superfluid regime (Fig.~\ref{fig3sites}). It is also illustrative to see how the exponent may change its value as the configuration equilibrates and initially empty or near-empty sites become near-full (Fig.~\ref{figanomdiff2mplus}). In accordance with the general rule, once the ratio of occupation numbers (actions) $I_n/I_{n-1}$ becomes large, one of the sites is effectively empty compared to the other and the exponent changes from $\zeta=2$ to $\zeta=4$.

\begin{figure}[H] 
\centering
\includegraphics[width=.9\linewidth]{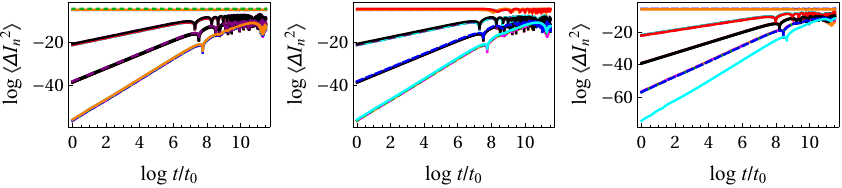}
\caption{Log-log plot of the second central moment (variance) of selected actions (it is unfeasible to plot 100 actions) for an ensemble of orbits in the chain of length $L=100$, with $U/J=1.25$ and $\mu=0$, with three different initial conditions -- initially filled sites are $n=30,60,90$ (left), $n=30,90,95$ (center) and $n=10,20,30,40,50,60$ (right). The phase space is still mixed, the same universal scaling laws apply as for short chains, and the calculations are still well-controlled.}
\label{figlong}
\end{figure}

We also want to emphasize that our findings (and our numerics) stay valid even for very long chains (Fig.~\ref{figlong}). In most of the paper we have kept $L=10$ just in order to be able to show full information in the figures (it is unfeasible to plot 100 curves) but there is no problem of computational nature to reach, e.g., $L=100$ (integration up to times $t/t_0\sim 10^6$ for an ensemble with 100 orbits takes less than half an hour of computer time even for $L=100$). Also, the anomalous scaling laws remain the same, and the phase space stays mixed: despite the general belief that mixed dynamics is mainly a characteristic of nonintegrable few-body systems, for classical Bose-Hubbard chains it is still present even with hundreds of degrees of freedom, despite the nonintegrability of the system. It would be interesting to check analytically if the regular islands persist even in the limit $L\to\infty$.

\begin{figure}[H] 
\centering
\includegraphics[width=.45\linewidth]{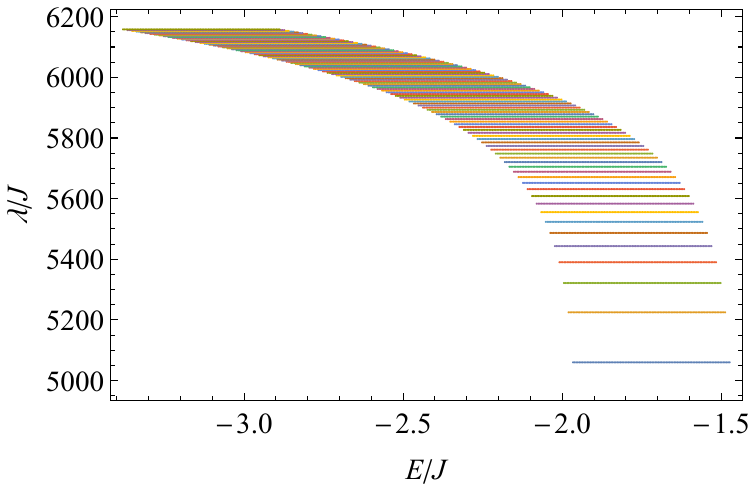}
\caption{Lyapunov exponent for a set of orbits with different $0<U/J<110$ and $0<\mu/J<2$, plotted as a function of the initial energy (determined by $U/J$, $\mu/J$ and the initial conditions). Chaos becomes weak for orbits with positive energies as they are dominated by the kinetic energy and thus are nearly regular. The near-horizontal lines of different colors correspond to the orbits with the same $U$ but different $\mu$: to a very good approximation, only $U$ determines chaos whereas mostly $\mu$ determines the energy.}
\label{figheatmap}
\end{figure}

Finally, in relation to the connection between transport and chaos, it might be useful to show the Lyapunov exponent not only as a function of $U,\mu$ and the initial conditions (although that is indeed our main concern) but also as a function of energy. This is the way the route to chaos is studied for classical Bose-Hubbard equations in \cite{PauschOptimalRoute,Nakerst:2022prc,Dahan:2022classical}. Fig.~\ref{figheatmap} shows the Lyapunov exponent for a set of orbits with different $\mu,U$ values tuned so that the orbits start with different energies. The roughly parabolic decay of $\lambda$ with the growth of energy captures one of the two branches of the dependence $\lambda(E)$ studied in \cite{Nakerst:2022prc}. The decay of chaos with increasing energies is intuitive, as highly energetic orbits are dominated by the kinetic energy and little influenced by interactions. We do not see the growing branch, obtained in \cite{Nakerst:2022prc} for larger negative energies (understood from the fact that orbits with very large negative energies are strongly localized and thus effectively see just a harmonic oscillator potential), probably because of different initial conditions and nonzero chemical potential (which was not considered in \cite{Nakerst:2022prc}). An important conclusion is that $\lambda$ is almost independent of $\mu$, therefore the frequent focus on the $\mu=0$ case in the literature is justified.
 
\section{\label{secappmaxle}Behavior of the maximum Lyapunov exponent}

In Figs.~\ref{figtable}-\ref{figtabtabledouble} we have given the behavior of Lyapunov exponents for individual $I_n$ variables. This is of primary interest for us as we want to understand the connection -- if any -- between the strength of chaos and the diffusion scaling exponent. We have found that indeed both chaos and anomalous diffusion depend on the initial condition -- filled vs. empty, though diffusion more crucially so. On the other hand, the dependence on $U/J$ and $\mu/J$, while almost non-existent for diffusion, is present also for the Lyapunov exponents -- they first rise (up to about $U/J\sim 5$) then fall with increasing $U/J$, where they rise with $\mu/J$ for the whole interval we have studied. 

\begin{figure}[H] 
\centering
\includegraphics[width=.9\linewidth]{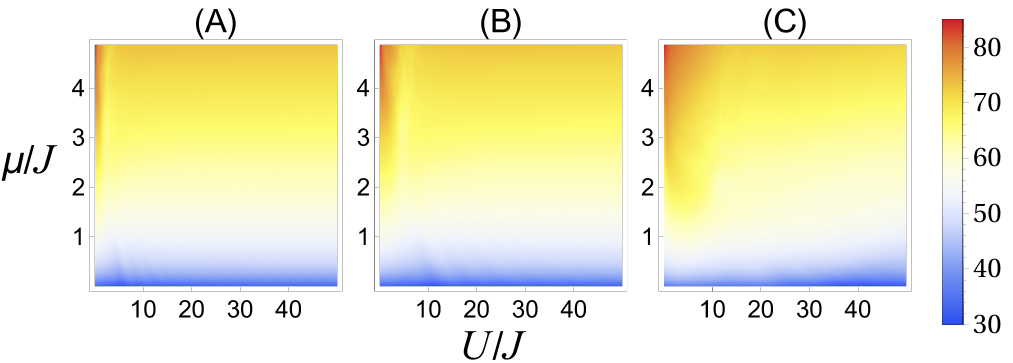}
\caption{The maximum Lyapunov exponent $\lambda_m$ (in units of $J$) as a function of the Coulomb interaction $U/J$ and the chemical potential $\mu/J$, for a Bose-Hubbard chain with $L=10$ wells, (A) with the well $n=5$ initially full and the rest empty (B) with the wells $n=4$ and $n=5$ initially full and the rest empty (C) with all wells initially full with random fillings (adding to $2$ in accordance with the constraint). The Coulomb coupling $U/J$ is varied from $0.5$ to $50.0$, i.e. we do not include the cases with $U/J=0$ for these have exactly zero Lyapunov exponent. The overall behavior of the maximum Lyapunov exponent is similar to the behavior of individual exponents in Figs.~\ref{figtable}-\ref{figtabtabledouble}, however fine details such as the optimal and non-optimal routes to chaos cannot be seen in the maximum exponent.}
\label{figtabtablemax}
\end{figure}

We can instead look at the maximum Lyapunov exponent, i.e. the exponent in the direction of maximal growth of the variation. While often used in the literature as the more robust measure of chaos, it washes away some interesting properties of local dynamics, in particular the optimal routes to chaos and order seen as bright and dark lines in Figs.~\ref{figtabtablesingle}-\ref{figtabtabledouble}. In Fig.~\ref{figtabtablemax} we plot the dependence of the maximum exponent $\lambda_m$ for the same range of parameter values. Overall dependence on interactions and chemical potential is the same, but much more homogeneous and without optimal and non-optimal routes. This is expected as the chaos i nthe fully developed regime should be less sensitive to initial conditions. However, for purposes of understanding transport, such an indicator is not very informative.

\section{\label{secapplang}Canonical form of the Langevin equation}

Here we attempt to transform the Langevin equation (\ref{langa}) into the canonical form, where the right-hand side does not depend on the variables $a_i$, instead it is in general the sum of a (possibly time-dependent) deterministic external force $\mathbf{F}$ and (possibly time-dependent) noise term $\hat{G}(t)d\mathbf{W}$, where the $\mathbf{W}$ is a Wiener process with variance $w^2$. We follow the procedure for the change of variables $\mathbf{a}\mapsto\mathbf{A}$ in Wiener processes \cite{CohenElliott:2015book} and adopt (for this Appendix) the vector notation, denoting the vectors $a_j,j=1,\ldots N$ by boldface letters (in addition to denoting matrix by a hat, as we did already in the main text):
\begin{eqnarray}
d\mathbf{a}(t)&=&\hat{g}(t,\mathbf{a})d\mathbf{W}\mapsto d\mathbf{A}(t)=\mathbf{F}(t)dt+\hat{G}(t)d\mathbf{W}\\
d\mathbf{A}&=&\left(\frac{\partial\mathbf{A}}{\partial t}+\frac{1}{2}w^2\hat{g}^2\cdot\frac{\partial^2\mathbf{A}}{\partial\mathbf{a}^2}\right)dt+\frac{\partial\mathbf{A}}{\partial\mathbf{a}}\cdot\hat{g}\cdot d\mathbf{W}\\
\mathbf{F}&=&\frac{\partial\mathbf{A}}{\partial t}+\frac{1}{2}w^2\hat{g}^2\cdot\frac{\partial^2\mathbf{A}}{\partial\mathbf{a}^2},~~\hat{G}=\hat{g}\cdot\frac{\partial\mathbf{A}}{\partial\mathbf{a}}\\
\frac{\partial\mathbf{A}}{\partial\mathbf{a}}&=&\hat{g}^{-1}\cdot\hat{G},~~\frac{\partial^2\mathbf{A}}{\partial\mathbf{a}^2}=-\hat{g}^{-2}\cdot\frac{\partial\hat{g}}{\partial\mathbf{a}}\cdot\hat{G}\\
\frac{\partial^2\mathbf{A}}{\partial\mathbf{a}\partial t}&=&\hat{g}^{-1}\cdot\frac{d\hat{G}}{dt}-\hat{g}^{-2}\frac{\partial\hat{g}}{\partial t}\cdot\hat{G}\\
\frac{\partial F}{\partial\mathbf{a}}&=&0\Rightarrow\frac{d\log\hat{G}}{dt}=\frac{\partial\log\hat{g}}{\partial t}+\frac{w^2}{2}\hat{g}\cdot\frac{\partial^2\hat{g}}{\partial\mathbf{a}^2}=0\\
0&=&\frac{\partial}{\partial\mathbf{a}}\left(\frac{\partial\log\hat{g}}{\partial t}+\frac{w^2}{2}\hat{g}\cdot\frac{\partial^2\hat{g}}{\partial\mathbf{a}^2}\right)~~\textrm{solvability condition}\label{langasolcond}\\
\hat{G}&=&\exp\left[\frac{w^2}{2}\int dt\hat{g}\cdot\frac{\partial\hat{g}}{\partial\mathbf{a}}\right]\\
\mathbf{F}&=&\frac{w^2}{2}\int d\mathbf{a}\left(\frac{\partial^2\hat{g}}{\partial\mathbf{a}^2}-\frac{\partial\hat{g}}{\partial\mathbf{a}}\right)\cdot\exp\left[\frac{w^2}{2}\int dt\hat{g}\cdot\frac{\partial\hat{g}}{\partial\mathbf{a}}\right]
\end{eqnarray}
For our system, since $\partial\hat{g}/\partial t=0$, the solvability condition (\ref{langasolcond}) boils down to $\hat{g}'\cdot\hat{g}''+\hat{g}\cdot\hat{g}'''=0$ yielding $\log(\hat{g}\cdot\hat{g}'')=\mathrm{const.}$ which means either $\hat{g}=\hat{\mathbf{c}}_0+\hat{c}_1\cdot\mathbf{a}$ for constant $\mathbf{c}_0,\hat{c}_1$ (if the constant on the right-hand side equals zero), or it implies a solution for $\hat{g}$ in terms of inverse error functions. The latter is manifestly not the case for our $\hat{g}$ and the former is only true if we ignore the constraint and thus indeed only have a linear dependence of $\hat{g}$ on $\mathbf{a}$. Then however the matrix $\hat{g}$ is not invertible hence there is no solution. If we include the constraint and solve the effective system with $L-1$ sites as in (\ref{langatilde}) then we do not satisfy the condition (\ref{langasolcond}), hence there is again no solution (or in other words, the expression for $\hat{\bar{G}}$ in the last line is not $\mathbf{\bar{a}}$-independent as we wanted it to be). We conclude that this system cannot be reduced to a canonical Langevin equation with $\mathbf{\bar{a}}$-independent right-hand side.

\section{\label{secappnlse}Anomalous transport versus dynamics}

Looking at the description of dynamics and Lyapunov exponents in Section \ref{secdyn}, we have concluded that the transport exponents mainly have to do with the distribution of occupation numbers, just like the Lyapunov exponents in Figs.~\ref{figtabtablesingle} and \ref{figtabtabledouble}: the existence of resonances and the strength of nonintegrability (i.e., $U/J$ ratio) both have little influence on the scaling laws, and likewise do not influence much the chaoticity of individual orbits. Rather, chaos is the characteristic of a near-filled site $n$ surrounded by near-empty sites $n-1,n+1$ (in the sense that $I_n>I_{n+1},I_{n-1}$), and this situation also implies the transport exponent $0$. The further we are from a filled site, the smaller the Lyapunov exponent and the larger the transport exponent. Therefore, large transport exponents (which are far from normal diffusion) indeed go hand in hand with weak chaos, as found in many circumstances in \cite{ZASLAVSKY2002461}. The long-term universal normal diffusive regime is reached only after the initial cell (ensemble) in phase space has spread out sufficiently to "forget" its initial condition.

\begin{figure}[H] 
\centering
\includegraphics[width=.9\linewidth]{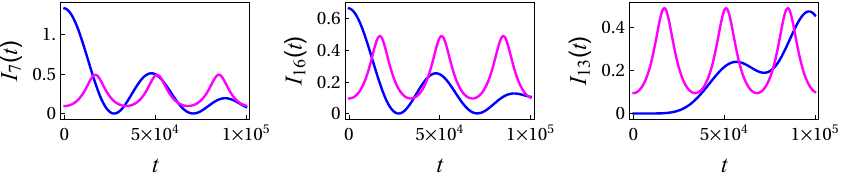}
\caption{Time evolution of the action for sites $7,13,16$, for a chain of length $L=20$, with $U/J=13.3$, with sites $n=7$ and $n=16$ initially filled, obtained from the numerics (blue) and from the $J=0$ DNSE (magenta). The period of oscillations is very well predicted by DNSE, although the amplitudes and the overall shape differ significantly, particularly for the initially empty sites (the rightmost panel). Nevertheless, we find even such very limited agreement unexpected for such a simple approximation. It suggests that early-time dynamics is collective in nature and thus possesses many additional near-integrals of motion; they cease to be near-conserved at longer times, when thermalization and anomalous diffusion kick in.}
\label{figdnse}
\end{figure}

For finite $\mu$ and large $U/J$ values, i.e. in the Mott-like regime, the discrete nonlinear Schr\"odinger equation which, as we have mentioned previously, follows from the continuum limit of the Bose-Hubbard chain, provides a simple but useful model of transport. The first step is to write the actions as $I_j(t)=a_j^*(t)a_j(t)$, with $a_j(t)=f_j(t)\exp\left(\imath\phi_j(t)\right)$, the constraint now becoming $\sum_j\vert f_j\vert^2=1$. Plugging this into the equations of motion, we get
\begin{equation}
\imath\dot{f}_j+J\left(f_{j-1}+f_{j+1}\right)-\mu f_j-U\vert f_j\vert^2f_j=0.
\end{equation}
The homogeneous ($J=0$) solution satisfying the initial condition $I_j(t=0)=I_0$ reads
\begin{equation}
f_j(t)=\frac{\sqrt{\mu}}{\sqrt{\left(\frac{\mu}{I_0}-U\right)e^{2\imath\mu t}+U}}.
\end{equation}
Squaring the module of the above solution we can obtain the actions. The zeroth order result in $J$ (i.e., $J=0$) is the only one we can obtain analytically. But a look at Fig.~\ref{figdnse} suggests it is not a bad approximation for high $U/J$ values, in particular the period $2\mu$ is indeed well matched to the numerical result. This result holds at shorter timescales; in the long-term, the integrable, zero-hopping approximation breaks down. The morale is that the short-time, anomalous transport regime is dominated by collective motion with many quasi-integrals of motion (for $J=0$ they become exact integrals of motion): the collective character of the DNSE leads to long-range correlations and anomalous diffusion. For longer times, the $J=0$ approximation is no longer meaningful and we lose the quasi-integrable character of dynamics, the long-distance correlations and anomalous diffusion, again in accordance with the intuition of \cite{ZASLAVSKY2002461,zaslavsky2007physics}.

\bibliography{bhRefs}
%\nolinenumbers
\end{document}